# N-, P-, As-triphenylene-graphdiyne: Strong and stable 2D semiconductors with outstanding capacities as anodes for Li-ion batteries

Bohayra Mortazavi[*,1], Masoud Shahrokhi[l,2], Mohamed E. Madjet[l,3], Meysam Makaremi[4], Said Ahzi[3,5] and Timon Rabczuk[6,#]

[1]Institute of Structural Mechanics, Bauhaus-Universität Weimar, Marienstr. 15, D-99423 Weimar, Germany.
[2]Institute of Chemical Research of Catalonia, ICIQ, The Barcelona Institute of Science and Technology, Av. Països Catalans 16, ES-43007 Tarragona, Spain.
[3]Qatar Environment and Energy Research Institute, Hamad Bin Khalifa University, Qatar Foundation, Doha, Qatar.
[4]Department of Materials Science and Engineering, University of Toronto, 184 College Street, Suite 140, Toronto, ON M5S 3E4, Canada.
[5]Université de Strasbourg, ICube–CNRS, Strasbourg, France
[6]Department of Computer Engineering, College of Computer and Information Sciences, King Saud University, Riyadh, Saudi Arabia.

**Abstract**

Since the first report of graphdiyne nanomembranes synthesis in 2010, different novel graphdiyne nanosheets have been fabricated. In a latest experimental advance, triphenylene-graphdiyne (TpG), a novel two-dimensional (2D) material was fabricated using the liquid/liquid interfacial method. In this study, we employed extensive first-principles simulations to investigate the mechanical/failure, thermal stability, electronic and optical properties of single-layer TpG. In addition, we predicted and explored the properties of nitrogenated-, phosphorated- and arsenicated-TpG monolayers. Our results reveal that TpG, N-TpG, P-TpG and As-TpG nanosheets can exhibit outstanding thermal stability. These nanomembranes moreover were found to yield linear elasticity with considerable tensile strengths. Notably, it was predicted that monolayer TpG, As-TpG, P-TpG and N-TpG show semiconducting electronic characters with direct band-gaps of 1.94 eV, 0.88 eV, 1.54 eV and 1.91 eV, respectively, along with highly attractive optical properties. We particularly analyzed the application prospect of these novel 2D materials as anodes for Li-ion batteries. Remarkably, P-TpG and N-TpG nanosheets were predicted to yield ultrahigh charge capacities of 1979 mAh/g and 2664 mAh/g, respectively, for Li-ions storage. The acquired results by this work suggest TpG based nanomembranes as highly promising candidates for the design of flexible nanoelectronics and energy storage devices.

Corresponding authors: *bohayra.mortazavi@gmail.com, #timon.rabczuk@uni-weimar.de;
[l]These authors contributed equally.

## 1. Introduction

Graphene [1,2] exceptional physics and chemistry, has emerged the two-dimensional (2D) materials as a novel class of materials. During the last decade, enormous experimental and theoretical efforts have been devoted to fabricate and predict new 2D materials and heterostructures. This class of materials includes a wide-variety of nanomembranes that have been experimentally realized, which can be categorized into different families. Graphene itself belongs to the monoelemental 2D family, which also includes other prominent members, like: silicene [3,4], germanene [5], stanene [6], phosphorene [7,8] and borophene [9,10]. For many applications, especially those in connection with nanoelectronics and optoelectronics, presenting a band-gap is a critical requirement. Among the aforementioned experimentally synthesized monoelemental 2D materials, only the phosphorene is well-understood to exhibit semiconducting electronic character and the other members show either metallic or zero band-gap electronic nature. Nevertheless, other families of 2D materials with inherent semiconducting electronic characters have garnered remarkable attentions. In this regard, 2D materials families of transition metal dichalcogenides, like $MoS_2$ and $WS_2$ [11,12], and carbon-nitride nanosheets, like graphitic carbon nitride g-$C_3N_4$ [13,14], nitrogenated holey graphene, $C_2N$ [15] and 2D polyaniline, $C_3N$ [16] have been experimentally realized and extensively studied in the literature.

A couple of decades before the rise of graphene [2], graphdiyne family, which includes full carbon 2D allotropes, with hybrid sp and $sp^2$ covalently bonded atoms arranged in various crystal lattices were theoretically predicted [17]. First-principles calculations notably confirm semiconducting electronic character for different graphdiyne family members [18–25], desirable for post-silicon nanoelectronics. Besides that, graphdiyne family members were also theoretically proven to exhibit outstanding properties, suitable for various applications, such as; anode materials for rechargeable batteries [26–29], hydrogen storage [30–33], catalysts [34] and thermoelectricity [35,36]. First experimental success with respect to synthesis of graphdiyne was reported in 2010 by Li *et al.* [37], via a cross-coupling reaction using the hexaethynylbenzene. In 2017, Jia *et al.* [38] and Matsuoka *et al.* [39] reported the experimental realization of two different graphdiyne structures. Jia *et al.* [38] reported the fabrication of carbon Ene-yne from tetraethynylethene by solvent-phase reaction and in the work by Matsuoka *et al.* [39] crystalline graphdiyne nanosheets were produced at a gas/liquid or liquid/liquid interface. Ground-breaking experimental advances bring the research on the graphdiyne family into a new level [18,37]. It is worthy to note that experimental observations [40–42] have confirmed that full carbon graphdiyne structures can exhibit outstanding prospects for critical technologies, such as the; energy storage systems [43–47], catalysts [44,48], water reduction [49] and electrochemical actuators [50]. Another novel and attractive concept with respect to the graphdiyne structures, is to design and fabricate novel lattices, in which some carbon atoms are replaced with other elements, in order to further enhance the efficiencies for particular applications. In this regard, using the experimental concept proposed by Matsuoka *et al.* [39], Kan *et al.* [51] recently succeeded in the synthesis of Nitrogen-graphdiyne nanomembranes. In another exciting recent experimental advance, Wang *et al.* [52] reported recently the synthesis of Boron-graphdiyne nanosheets through a bottom-to-up synthetic strategy. Our recent theoretical studies concerning the Nitrogen-graphdiyne [53] and Boron-graphdiyne [54] monolayers reveal their semiconducting electronic character, outstandingly high stretchability, low thermal conductivity and very promising optical properties. According to our first-principles results [54], B-graphdiyne was astonishingly predicted to yield ultrahigh charge capacities of 5174 mAh/g and 3557 mAh/g for Ca- and Li-ions storage, respectively.

As it is clear, the research on the graphdiyne 2D materials family is getting more attractive, originated not only from the latest experimental advances but also due to their highly promising application prospects for critical technologies like, energy storage, catalysis, stretchable nanoelectronics, nanosensors and biomedicine [55–58]. In a recent experimental study by Matsuoka *et al.* [59], a novel graphdiyne structure, so called triphenylene graphdiyne (TpG) was designed and fabricated experimentally. TpG nanomembranes were synthesized employing a modified liquid/liquid interfacial synthesis method [39] using the hexaethynyltriphenylene monomer [59]. This latest experimental success in the fabrication of TpG nanomembranes raises the importance of theoretical studies in order to provide understanding of the intrinsic material properties [53,54,60,61]. Motivated by this experimental advance, we also predicted nitrogenated-TpG (N-TpG), phosphorated-TpG (P-TpG) and arsenicated-TpG (As-TpG) nanomembranes. We then employed first-principles density functional theory (DFT) simulations to explore the mechanical, thermal stability, optical and electronic properties of these novel 2D systems. Because of the fact that 2D materials have currently garnered remarkable attention for the application as anode materials for Li-ion batteries [62–69], we also employed the DFT calculations to explore the possible application of TpG, As-TpG, P-TpG and N-TpG nanosheets as anode materials for Li-ions storage. The acquired results suggest very promising performances of this novel class of 2D materials and will hopefully motivate further theoretical and experimental studies.

## 2. Computational methods

First-principles calculations in this work were conducted using the Vienna *Ab initio* Simulation Package (VASP) [70–72] within the Perdew-Burke-Ernzerhof (PBE) functional [73] for the exchange correlation potential. The core electrons are replaced by Projected augmented wave (PAW) method [74] and pseudo-potential approach. A plane-wave cutoff energy of 500 eV was set for the DFT calculations. We used VESTA [75] package for the visualization of atomic structures and charge distributions. In order to acquire the energy minimized structures, the size of the unit-cell was uniformly changed and then the conjugate gradient method was employed for the geometry optimizations using a 7×7×1 Monkhorst-Pack [76] k-point mesh size. In this case, criteria for the convergence of the energy of electronic self consistence-loop was considered to be $10^{-4}$ eV and for the structural relaxation, the Hellmann–Feynman forces on each atom was taken to be 0.01 eV/Å. To evaluate the electronic band-structure, we used a denser k-point mesh size of 11×11×1. Because of the fact that the PBE underestimates the band-gap values of semiconductors, we also employed the screened hybrid functional, HSE06 [77] with a 5×5×1 k-point mesh size to provide more accurate estimations. To calculate the optical properties, 12×12×1 k-point grids have been used.

Uniaxial tensile simulations were conducted in order to evaluate the mechanical properties. To this aim, the periodic simulation box size along the loading direction was increased gradually with a fixed engineering strain increment of 0.001. To ensure the uniaxial stress-conditions, the simulation box size along the sheet perpendicular direction of loading was altered to also reach a negligible stress [53,54]. We then used conjugate gradient method to simulate the atomic rearrangements during the uniaxial tensile simulations, with the same termination criteria as those we utilized to reach the energy minimized structures, however, with a coarser k-point mesh size of 5×5×1 to enhance the calculations speed. In order to examine the thermal stability of considered structures, ab initio molecular dynamics (AIMD) simulations were performed for the hexagonal unit-cells, using the Langevin thermostat with a time step of 1 fs and 2×2×1 k-point mesh size [53,54].

For the modelling of Li atoms adsorption over the unit-cell structures of TpG, As-TpG, P-TpG and N-TpG monolayers, a dispersion scheme of DFT-D2 [78] was selected to improve the binding energy calculations by accounting for the dispersion corrections. To find energy minimized structures, the conjugate gradient method was employed with the convergence criteria of 0.01 eV/Å for the forces, using 3×3×1 k-point girds. In order to more precisely report the final energy values and charge densities, we conducted single point calculations over the energy minimized structures, using the tetrahedron method with Blöchl corrections [79] with a 5×5×1 k-point mesh size [54]. Bader charge analysis [80] was conducted to evaluate the efficiency in the charge transfer from the Li adatoms to the monolayers. The climbing image nudged elastic band (CiNEB) method [81] was utilized in order to estimate energy barriers for the diffusion of a single Li adatom.

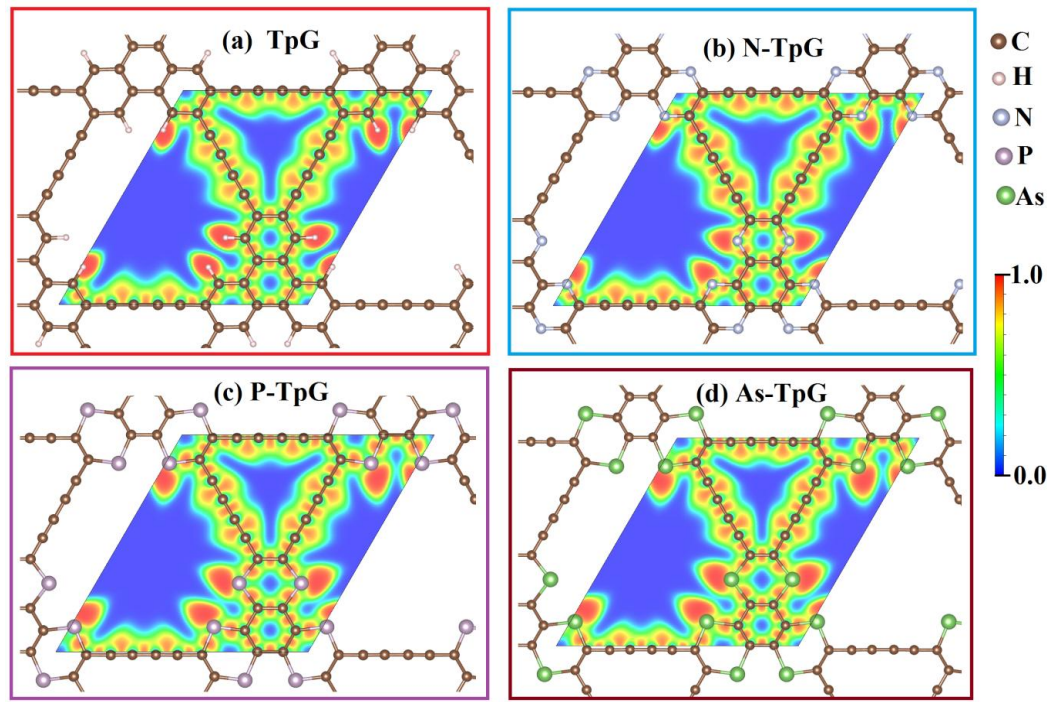

Fig. 1, Atomic structure of single-layer TpG, As-TpG, P-TpG and N-TpG. Contours illustrate the electron localization function [82] within the unit-cell. Horizontal and vertical directions correspond to the armchair and zigzag directions.

## 3. Results and discussions

In Fig.1, energy minimized TpG, As-TpG, P-TpG and N-TpG monolayers with hexagonal atomic lattices are illustrated. We should remind that for the doping of the original TpG nanosheet with other elements, numerous configurations can be chosen. In this work in order to predict more realistic structures, resembling experimentally characterized carbon-nitride 2D materials, like; N-graphdiyne [51], s- and tri-triazine-based graphitic carbon nitrides [13,14] and nitrogenated holey graphene [15], we only considered the doping of the $sp^2$ hybridized carbon atoms in the original hexagonal carbon rings of TpG. The hexagonal lattice constants of energy minimized TpG, As-TpG, P-TpG and N-TpG were acquired to be 13.783 Å, 14.767 Å, 14.415 Å and 13.496 Å, respectively. Unit-cells of energy minimized structures are given in the supporting information (SI) document. The HC-C bond length in the TpG lattice was found to be 1.40 Å. The average C-N, C-P and C-As bond lengths were measured to be 1.34 Å, 1.76 Å and 1.91 Å, respectively. As it is clear, N-TpG monolayer exhibits the closet lattice to the pristine TpG, with a very uniform bond lengths, whereas in the cases of P-TpG and As-TpG considerable bond elongations are observable around the P and As atoms. The energy per atoms of monolayer TpG, N-TpG, P-TpG and As-TpG were also calculated to be, -7.861 eV, -8.459 eV, -7.763 eV and -7.508 eV, respectively. These results confirm the higher energetic stability of N-TpG in comparison with P-TpG and As-TpG. To provide more insight concerning the bonding nature in the studied 2D structures, in Fig. 1 we also plotted the electron localization function (ELF) [82], which is a position-dependent function that takes a value from 0 to 1. The ELF equals to one, zero and half correspond to the regions with completely electron localizations, completely delocalized electron and electron gas-like behaviour [82], respectively. According to the ELF contours illustrated in Fig. 1, ELF values around the center of all bonds in the considered monolayers are greater than 0.8, confirming the dominance of covalent bonding in these novel nanosheets. Moreover, electron localization also occurs around the H, P, N and As atoms.

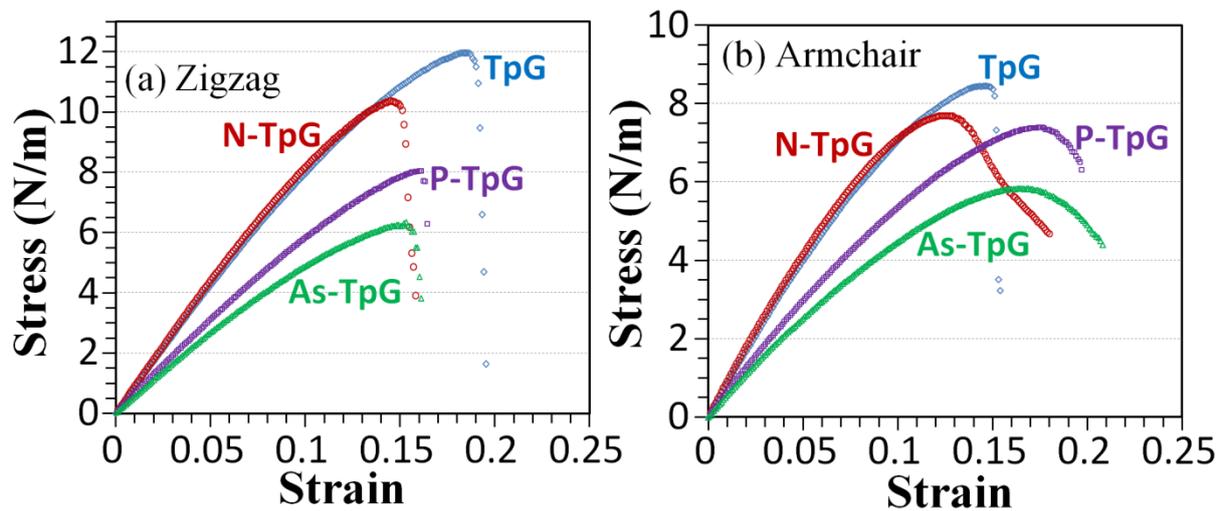

Fig. 2, Uniaxial stress-strain responses of single-layer TpG, As-TpG, P-TpG and N-TpG elongated along the (a) zigzag and (b) armchair directions.

### 3.1 Mechanical response

We first investigate mechanical properties of this novel class of 2D materials. In order to analyze the anisotropicity in the mechanical response, we conducted the uniaxial tensile simulations along the armchair and zigzag directions, in analogy to graphene. In Fig. 2, the DFT results for the uniaxial stress-strain responses of single-layer TpG, N-TpG, P-TpG and As-TpG elongated along the zigzag (Fig. 2a) and armchair (Fig. 2b) directions are compared. The predicted uniaxial stress-strain curves show initial linear responses corresponding to the linear elasticity. This is an interesting finding, since B-graphdiyne and N-graphdiyne [53,54] nanosheets were normally found not to yield linear elasticity (elastic instability), as at the early stages of loading their deformation are mainly achieved by structural deflection rather than the bond-elongation. Slope of the initial linear relation in the stress-strain curve can be used to obtain the elastic modulus. Within this range, the strain along the traverse direction of loading ($\varepsilon_t$) with respect to the loading strain ($\varepsilon_l$) is constant and can be used to evaluate the Poisson's ratio, using: $-\varepsilon_t/\varepsilon_l$ [83]. Another critical mechanical characteristics of a material are the ultimate tensile strength and its corresponding strain, which defines the stretchability of a material. In Table 1, the mechanical properties of TpG, N-TpG, P-TpG and As-TpG nanosheets are summarized. These results reveal that for every structure, the elastic modulus and Poisson's ratio for the loading along the armchair and zigzag are very close, and thus suggesting the convincingly isotropic elastic response. On the other side, along the zigzag these novel 2D systems exhibit considerably higher tensile strengths, confirming anisotropic tensile response. The ratio of anisotropy in tensile strength was found to decrease from TpG to N-TpG to P-TpG and to As-TpG. As it is clear, among the studied materials TpG and N-TpG nanosheets exhibit the highest tensile strength and elastic modulus, respectively. Interestingly, despite of the highly porous atomic lattices, the elastic modulus of TpG, N-TpG and P-TpG are considerably higher than some other densely packed 2D materials, like silicene, germanene and stanene [84].

Table 1, Summarized mechanical properties of single-layer TpG, N-TpG, P-TpG and As-TpG for the uniaxial loading along the armchair and zigzag directions. $E$, $P$, $UTS$ and $SUTS$ stand for elastic modulus, Poisson's ratio, ultimate tensile strength and strain at ultimate tensile strength point, respectively.

| Structure | Direction | $E$ (N/m) | $P$ | $UTS$ (N/m) | $SUTS$ |
|---|---|---|---|---|---|
| TpG | Armchair | 87.9 | 0.34 | 8.47 | 0.14 |
| | Zigzag | 87.8 | 0.34 | 11.95 | 0.18 |
| N-TpG | Armchair | 92.5 | 0.31 | 7.70 | 0.12 |
| | Zigzag | 93.4 | 0.31 | 10.37 | 0.14 |
| P-TpG | Armchair | 64.6 | 0.35 | 7.40 | 0.17 |
| | Zigzag | 65.3 | 0.35 | 8.04 | 0.16 |
| As-TpG | Armchair | 56.5 | 0.34 | 5.84 | 0.16 |
| | Zigzag | 55.4 | 0.31 | 6.34 | 0.15 |

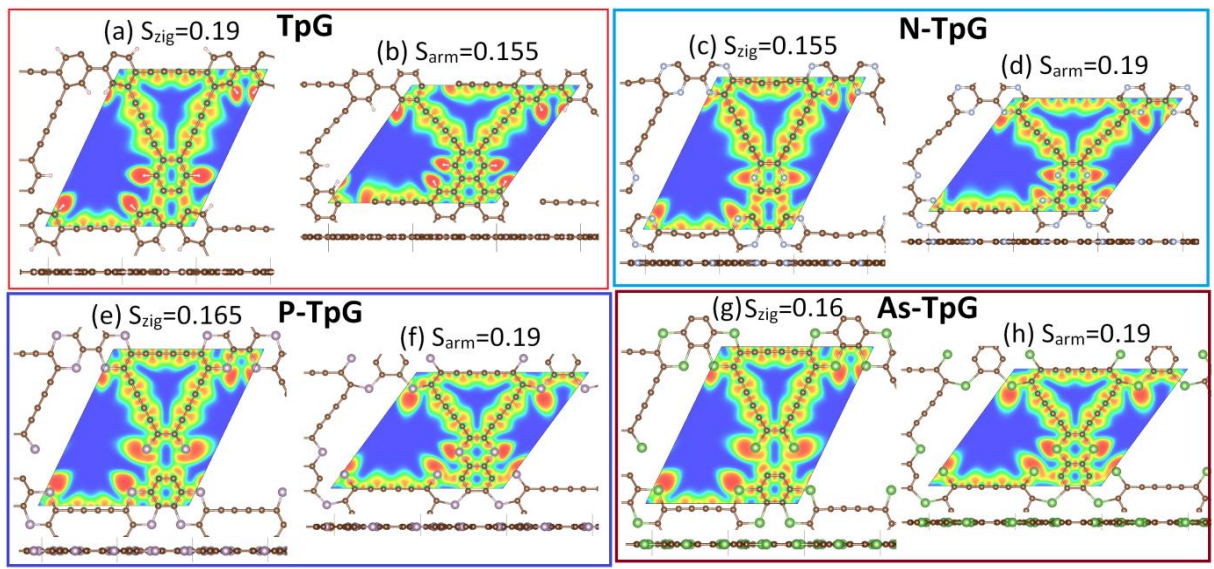

Fig. 3, Top and side views of the single-layer TpG, N-TpG, P-TpG and As-TpG at the strain levels close the ultimate tensile strength point. $S_{arm}$ and $S_{zig}$ depict the strain levels for the uniaxial loading along the armchair and zigzag directions, respectively. Contours illustrate the electron localization function [82].

To better understand the failure behaviour as well as underlying mechanisms resulting in higher tensile strengths of these monolayers along the zigzag direction, we consider the deformation process. In these systems, in a unit-cell exists three six-membered carbon chains, one exactly inline of armchair direction and the two oblique to the zigzag direction with an angle of 30 degree. As a general finding, during the uniaxial loading, bonds oriented along the loading direction stretch, whereas as a results of the sheet shrinkage along the width, other bonds oriented perpendicular of the loading contract. For the uniaxial loading along the armchair direction, from every three carbon-carbon chains existing in these monolayers, only one chain is exactly oriented along the loading direction and actively engages in the stretching and load bearing. On the other hand, for the uniaxial loading along the zigzag direction, from every three carbon chains, two involve in the load bearing and thus resulting in higher tensile strengths. The results for the side views shown in Fig. 3 reveal that during the uniaxial loading the planarity of the structures were kept intact as no out-of-plane movement is observable. Worthy to remind that in the case of B-graphdiyne [54], out-of-plane movement of atoms could result in superstretchability. We remind that rupture of bonds can be realized either by the remarkable increase in the bond length as compared with the other bonds in the system or by considering the ELF contours, where they extend and finally split into two parts. According to the acquired results illustrated in Fig. 3, the failure mechanism in TpG and N-TpG are similar. In these cases, for the uniaxial loading along the armchair the first rupture occurs in the $sp^2$-$sp^1$ carbon bonds. For the loading along the zigzag direction, the rupture in the TpG and N-TpG happens for the $sp^2$-$sp^2$ carbon bonds in the central full-carbon hexagonal ring. It can be concluded that the incorporation of N atoms in the original TpG slightly increases the rigidity, understandable from the enhancement of the elastic modulus, but it also yield weakening effect as it softens the $sp^2$-$sp^1$ and $sp^2$-$sp^2$ carbon bonds, leading to decline of the tensile strength. On the side, P-TpG and As-TpG demonstrate similar failure mechanism. For these nanosheets, for the loading along the armchair and zigzag the rupture initiates by the breakage of the P-C or As-C bonds. These bonds as discussed earlier, show considerable elongation in comparison with the C-C bonds. Interestingly, the stretchability of P-TpG and As-TpG are very close, however the lower rigidity of As-C bonds in comparison with P-C bonds is apparent as the tensile strength of As-TpG is distinctly lower than P-TpG.

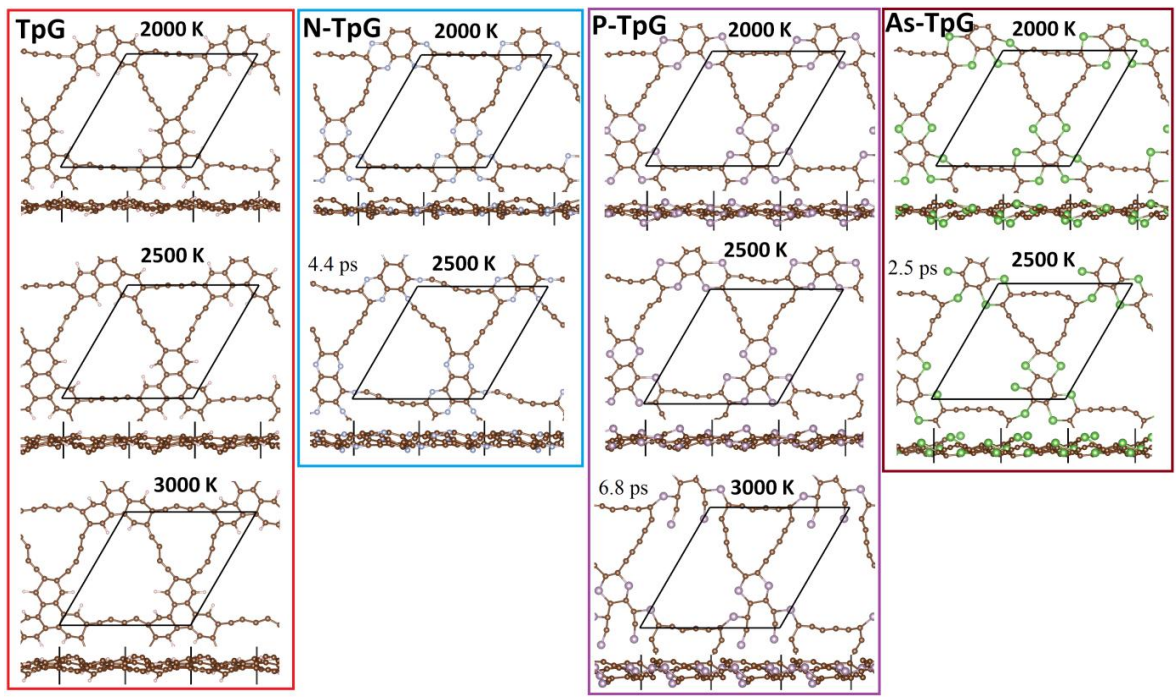

Fig. 4, Snapshots of the single-layer TpG, N-TpG, P-TpG and As-TpG at different temperatures. AIMD simulations were carried out for 20 ps and for every monolayer approximate simulation times that the ruptures initiate are mentioned explicitly.

## 3.2 Thermal stability

Despite of the fact that TpG, N-TpG, P-TpG and As-TpG nanomembranes exhibit linear elasticity and remarkable tensile strengths, these may not guarantee their thermal stability. To investigate the thermal stability of these novel 2D systems, we conducted the AIMD calculations at different temperatures, from 500 K to 3000 K with a 500 K temperature step. The simulations were performed for 20 ps and top and side views snapshots of the studied nanomembranes are depicted in Fig. 4. Interestingly, as it can be concluded from the results shown in Fig. 4, all considered nanosheets could stay intact at the extremely high temperature of 2000 K, which confirm their outstanding thermal stability. The original TpG monolayer could exhibit the highest thermal stability, with a perfectly kept atomic lattice up to 3000 K. P-TpG nanomembrane shows the second highest thermal stability, which can endure up to 2500 K. In this structure the first rupture was found to occur in the $sp^2$-$sp^2$ carbon bonds in the central full-carbon hexagonal ring. Both of N-TpG and As-TpG were however found to disintegrate at 2500 K. In the N-TpG nanosheet the failure initiates by the breakage of $sp^2$-$sp^2$ carbon bonds in the $C_4N_2$ rings, whereas in the As-TpG monolayer the lattice destroy first by the failure of As-C bonds. The AIMD results reveal outstanding thermal stability of this novel class of graphdiyne lattices. These results nonetheless cannot ensure dynamical stability of TpG, N-TpG, P-TpG and As-TpG nanomembranes, which should be explored in a separate investigation through analysis of phonon spectrum.

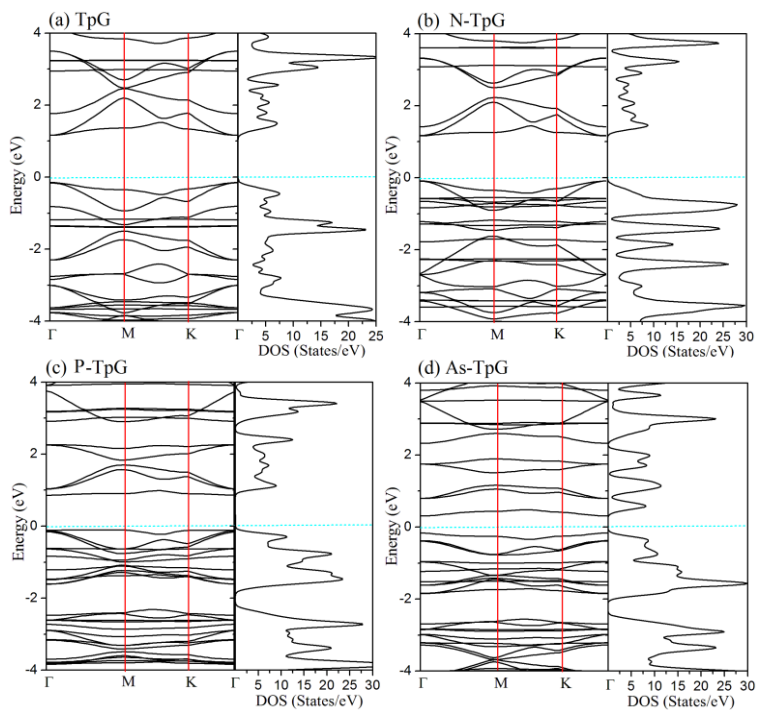

Fig. 5, Electronic band structure and total EDOS of single-layer TpG, N-TpG, P-TpG and As-TpG predicted by the PBE functional. The Fermi energy is aligned to zero.

## 3.3 Electronic and optical properties

To probe the electronic properties of single-layer TpG, N-TpG, P-TpG and As-TpG, electronic band structures were investigated. Fig. 5 illustrates the band structure along the high symmetry *Γ-M-K-Γ* directions and total electronic density of states (EDOS) predicted by the PBE method. Obtained results show that the both valence band maximum (VBM) and the conduction band minimum (CBM) of all monolayers occur at the Γ-point, resulting in direct band-gaps. Presenting a direct and narrow band-gap is highly promising for the applications in post silicon nanoelectronics. According to the PBE results the band-gaps of TpG, As-TpG, N-TpG and P-TpG monolayers are predicted to be 1.26 eV, 0.47 eV, 1.24 eV and 1.00 eV, respectively. Since the PBE functional underestimates the band-gap values, total EDOS of single-layer TpG, N-TpG, P-TpG and As-TpG were also calculated using the HSE06 hybrid functional (obtained results are shown in Fig. S1 of SI document). The band-gap values on the basis of HSE06 functional for the TpG, As-TpG, N-TpG and P-TpG monolayers were measured to be 1.94 eV, 0.88 eV, 1.91 eV and 1.54 eV, respectively. Table 2 summarizes the electronic band-gap of aforementioned nanosheets within the PBE and HSE06 functionals. These results suggest that the phosphorated- and arsenicated-TpG monolayers can lower the electronics band-gap and suit them better for the application as nanotransistors, while nitrogenating of TpG shows limited effect on the electronic band-gap. Low and direct band-gaps may also suggest the considered nanosheets for the thermoelectric applications, provided that the thermal conductivity of these materials are low. This way, investigation of the thermal conductivity of these nanosheets by the classical molecular dynamics simulations can be an important topic for the future studies [85–87].

Table 2. The electronic band-gap, the first peak of *Im ε,* the static dielectric constant and the first adsorption peak of single-layer TpG, As-TpG, N-TpG and P-TpG.

| Structure | Band-gap (eV) | | First peak of *Im ε* (eV) | Static dielectric constant | First absorption peak |
|---|---|---|---|---|---|
| | PBE | HSE06 | | | |
| TpG | 1.26 | 1.94 | 1.75 | 2.99 | 1.88 |
| As-TpG | 0.47 | 0.88 | 0.73 | 4.94 | 0.81 |
| N-TpG | 1.24 | 1.91 | 1.62 | 3.10 | 1.72 |
| P-TpG | 1.00 | 1.54 | 1.15 | 4.00 | 1.38 |

We next discuss the optical responses of single-layer TpG, As-TpG, N-TpG and P-TpG. In this case, we also compared the optical properties of these monolayers with those of the pristine graphene [53,88]. Optical properties, including the imaginary and real parts of dielectric and absorption coefficients were calculated on the basis of random phase approximation (RPA) method [89]. Because of the huge depolarization effect in the 2D planar geometry for light polarization perpendicular to the plane (out-of-plane) [90], we only focus on the optical absorption spectrum for light polarization parallel to the plane (in-plane). Due to the symmetric structure of these nanosheets, the optical spectra are isotropic for light polarizations along the x-axis (E||x) and y-axis (E||y), hence, only the optical properties for light polarizations along E||x are reported. Optical properties were described by photon frequency dependent dielectric function, $\varepsilon(\omega) = \text{Re}\,\varepsilon_{\alpha\beta}(\omega) + i\,\text{Im}\,\varepsilon_{\alpha\beta}(\omega)$, which can be acquired from the electronic structures (the details of optical calculations are the same as those we conducted in our recent study [53]). The imaginary and real parts of the dielectric function of TpG, As-TpG, N-TpG and P-TpG monolayers for in-plane direction obtained from RPA+PBE are illustrated in Fig. 6. The first peak of *Im ε* for TpG and N-TpG monolayers occur at 1.75 eV and 1.62 eV respectively, which are in the visible range and highly desirable for the practical applications in optoelectronic devices operating in visible spectral range. The corresponding values for As-TpG and P-TpG were found to be 0.73 eV and 1.15 eV, respectively, which are in IR and near IR (NIR) range of light, respectively. The value of the static dielectric constant (the real part of the dielectric constant at zero energy,

Re $\varepsilon_0$) for the TpG, As-TpG, N-TpG and P-TpG monolayers are 2.99, 4.94, 3.10 and 4.00, respectively. Table 1 summarizes the first adsorption peak of *Im ε* and the static dielectric constant of aforementioned structures.

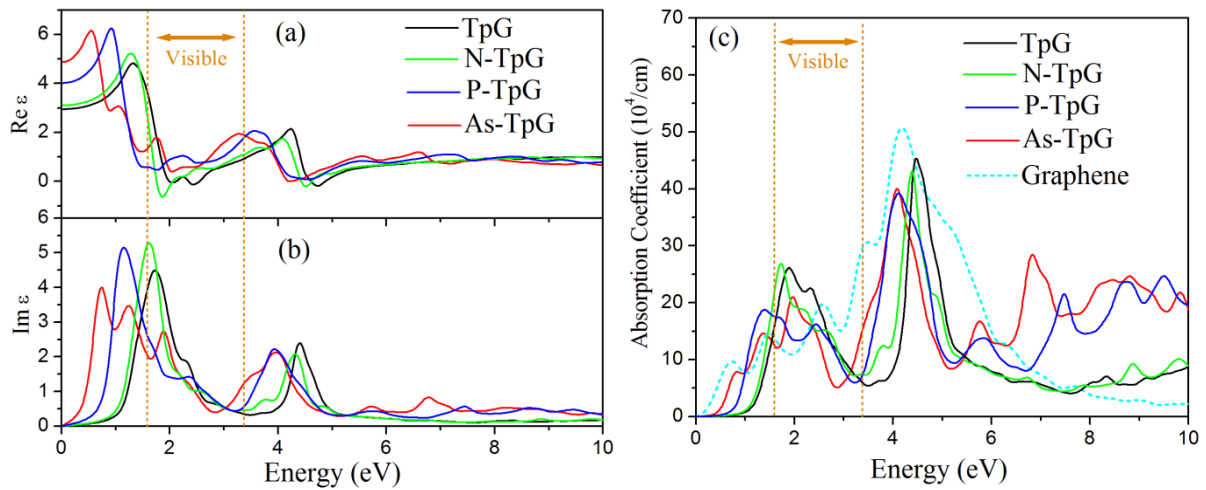

Fig. 6, (a) Imaginary and (b) real parts of the dielectric function and (c) absorption coefficient of single-layer TpG, As-TpG, N-TpG and P-TpG for the in-plane light polarizations, predicted using the PBE plus RPA approach. The cyan dashed line shows the in-plane absorption spectra for pristine graphene taken from our previous studies [53,88]. The visible range of the spectrum is shown by the orange dashed lines.

The absorption coefficient $\alpha_{\alpha\beta}$ for all considered monolayers are plotted in Fig. 6c. In this case we also compared acquired results with the absorption coefficient of pristine graphene derived from our previous studies [88]. The acquired results show that the first absorption peak for TpG and N-TpG nanosheet occur at visible range, 1.88 eV and 1.72 eV, respectively, while for As-TpG and P-TpG happen at 0.81 eV and 1.38 eV, respectively, which are in IR and NIR range of light, respectively. Moreover, it is clear that the absorption coefficient of aforementioned monolayers in energy range between 1.60-2.50 eV (from red to green) of visible light is larger than that of the graphene, whilst for frequency regime greater than 2.50 eV (from blue to violet) the absorption coefficient of graphene is larger than these nanomembranes. These findings confirm that the optical absorption of TpG monolayer in the range of red to green for visible light is better than that of the graphene, which is promising for the applications in optoelectronics and nanoelectronics. In general, the first absorption peaks reveal that these novel 2D nanostructures can absorb the visible, IR and NIR light, highlighting their promising prospect for applications in optoelectronics and nanoelectronics. It is worthy to mention that in this study we only considered the single-layer sheets. Since the optical and electronic properties may depend on the number of atomic-layers, investigation of thickness effect on the optical and electronic properties can be an attractive topic for the future studies.

### 3.4 Application as anode materials for Li-ion batteries

Because of the critical importance of the 2D materials in the design of advanced rechargeable metal-ion batteries, in the rest of this study we particularly explore the application prospect of TpG, N-TpG, P-TpG and As-TpG as anode materials for Li-ion batteries. To this aim, we first find the strongest binding sites and corresponding adsorption energies. The adsorption energy, $E_{ad}$, in this work was defined by [54]:

$$E_{ad} = E_{TLi} - E_T - E_{Li} \quad (1)$$

where $E_T$ is the total energy of pristine monolayers before Li atoms adsorption, $E_{TLi}$ is the total energy of the structures after Li atoms adsorption and $E_{Li}$ is the per atom energy of Li in the BCC lattice. The maximum adsorption energies for single Li atom over single-layer TpG, N-TpG, P-TpG and As-TpG were measured to be -1.83 eV, -2.83 eV, -2.21 eV and -2.58 eV, respectively. These results confirm that a single Li atom strongly interact and adsorb over the considered monolayers. As it is clear, N-TpG and TpG monolayers were found to yield the strongest and weakest adsorption with a single Li atom. In Fig. 7, first and second most favourable adsorption sites for a single Li atom over TpG, N-TpG, P-TpG and As-TpG along with the corresponding adsorption energies are illustrated. Interestingly, for all monolayers the strongest binding site occurs within the plane of atomic lattices. As shown in Fig. 7, in the cases of TpG, P-TpG and As-TpG monolayers, aforementioned site is within the triangular-like hollow site surrounded by 6 membered carbon chains. This hollow site nonetheless shows the second strongest adsorption energy for a Li atom over single-layer N-TpG. We also conducted the charge analysis to assess the charge transfer efficiency from a single Li atom to the considered monolayers and the acquired results are also shown in Fig. 7. Bader charge analysis [80] results confirm that from a single Li adatom the charge transfer efficiencies to the considered monolayers are always over 99%. These preliminary findings suggest that a single Li atom not only strongly adsorb over the single-layer TpG, N-TpG, P-TpG and As-TpG, but also becomes fully ionized upon the adsorption.

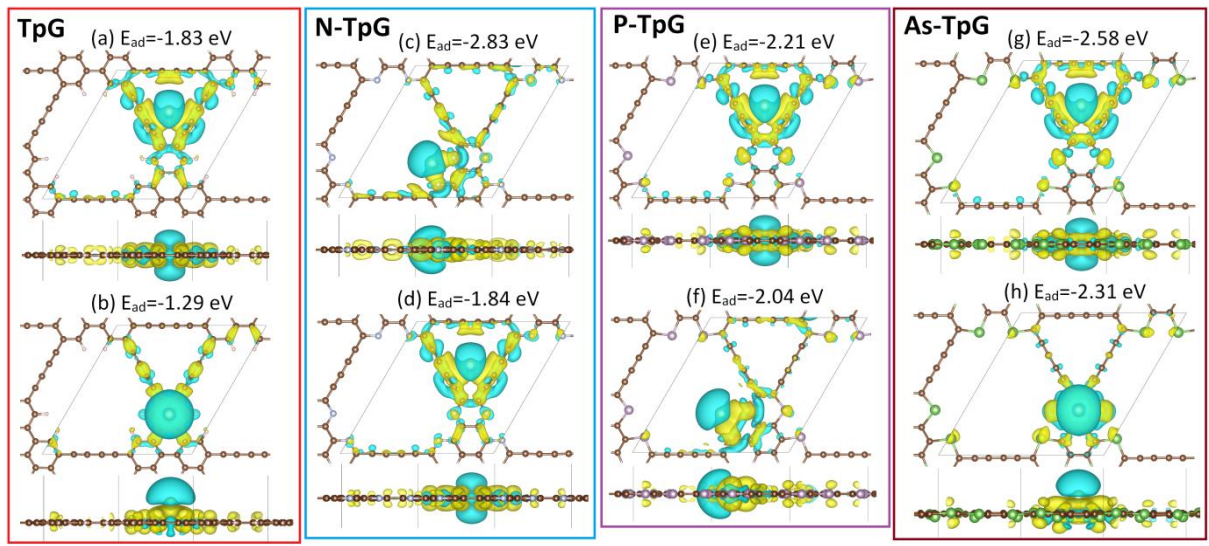

Fig. 7, Top and side views of the first and second most favourable adsorption sites for a single Li atom over TpG, N-TpG, P-TpG and As-TpG monolayers, respectively. Colour coding illustrate the binding charge transfer due to the adsorption of Li over the nanosheets, in which light-blue shows the charge losses and yellow reveals the charge gains.

The average adsorption energy profile as a function of the Li adatoms coverage strongly contributes to the charge capacity of a material as an anode for metal-ion batteries. Negative adsorption energy is required for the storage of metal adatoms. Moreover, increasing of Li atoms coverage may change or even degrade an anode material and subsequently affect the performance. To investigate these critical factors, we gradually and uniformly increased the Li adatoms coverage, first by filling the all strongest binding sites predicted earlier. Because of that fact that the possibilities for positioning the Li atoms over the studied monolayers are too numerous, for various Li atoms contents, three different structures were constructed and after the energy minimization, the results for the system with lowest energy was chosen for the average adsorption energy calculation, which was evaluated using the following equation:

$$E_{\text{av-ad}} = \frac{(E_{\text{TLi}} - n \times E_{Li} - E_T)}{n} \quad (2)$$

where $E_{TLi}$ is the total energy of monolayers with "$n$" adsorbed Li atoms. In Fig. 8a the evolution of average adsorption energy for Li-ions storage as a function of charge capacity over the single-layer TpG, N-TpG, P-TpG and As-TpG are illustrated. Worthy to remind that charge capacity is directly proportional to the adsorbed Li atoms ($n$) and can be obtained using the $nF/W_{sl}$ relation, where $F$ is the Faraday constant and $W_{sl}$ is the atomic mass of considered monolayers unit-cells. As a first finding consistent for all studied monolayers, by increasing of Li atoms content, average adsorption energy first sharply decreases. In the case of TpG, it is observable that after the adsorption of 15 Li atoms over the unit-cell lattice, equivalent with a charge capacity of 1097 mAh/g, the average adsorption energy reaches to a value very close to zero. For the As-TpG monolayers, despite of the acceptably negative adsorption energies up to the charge capacity of 472 mAh/g, we noticed that the adsorption of more Li atoms result in the considerable deflection of atomic lattice and thus we decided not to further consider this monolayer. According to the results shown in Fig. 8a, N-TpG and P-TpG exhibit highly promising and very similar adsorption energy relations, showing initial sharp drops and then reaching plateaus around the capacity of ~800 mAh/g, in which by further increasing of Li adatoms content the adsorption energy continues to decrease, but very smoothly. These observations reveal outstanding prospects of N-TpG and P-TpG nanomembranes for the application as anode materials in Li-ion batteries.

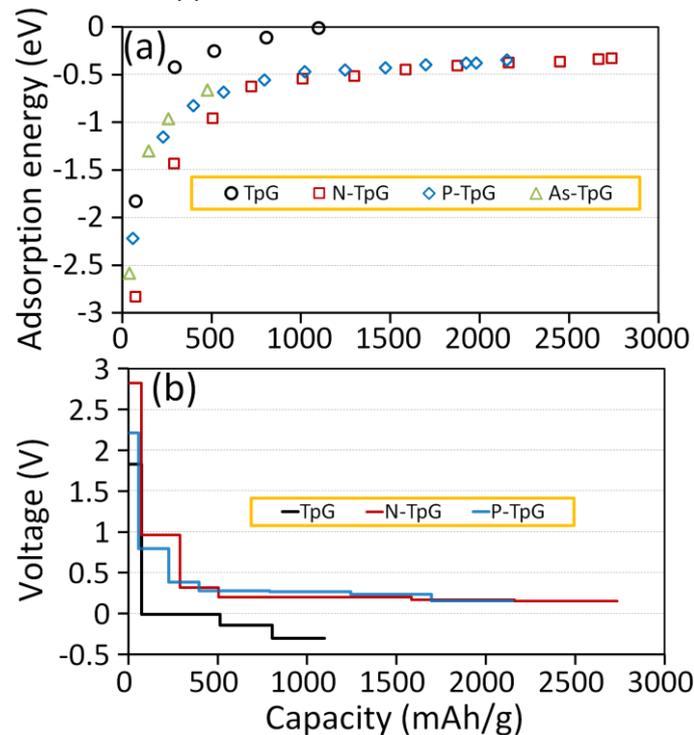

Fig. 8, (a) Average absorption energy and (b) open circuit voltage profiles as a function of charge capacity for Li-ions storage.

Appropriate open-circuit voltage profile is another critical requirement for the application of a material as an anode in rechargeable metal-ion batteries. Worthy to note that in a battery cell, output voltage is the difference in the cathode and anode solid phase potentials minus ohmic resistances at current collectors. In this way, an anode material with lower voltage profile can result in higher cell output voltage. Nonetheless, the voltage values close to zero can lead to undesirable effects, as it may for-example induce dendrite formation and leading to thermal runaway or other serious damages. As an approximation, the average voltage within the Li atoms coverage range of $x_1 \leq x \leq x_2$ can be predicted using the following equation [91,92]:

$$V \approx \frac{(E_{TLi_{x_1}} - E_{TLi_{x_2}} + (x_2-x_1)E_{Li})}{(x_2-x_1)} \quad (3)$$

where $E_{TLi_{x_1}}$ and $E_{TLi_{x_2}}$ are the total energies of the systems with $x_1$ and $x_2$ adsorbed Li adatoms, respectively [54]. In Fig. 8b predicted open circuit voltage profiles of TpG, N-TpG and P-TpG monolayer as a function charge capacity for Li-ions storage are plotted. As it is clear, in the case of pristine TpG, voltage value drops fairly early to a value close to zero. Worthy to note that in real applications, after first cycles of anode service, due to the formation of solid/electrolyte interface (SEI) layer the voltage curve may drop. Since the acquired voltage curves according to Eq. 3 are just approximations, it is very difficult to report the charge capacity of TpG nanosheets for Li-ions storage. On the other side, both N-TpG and P-TpG nanosheets exhibit very attractive voltage profiles, with values around ~0.3 V after the charge capacity of ~400 mAh/g. The voltage curves for aforementioned nanomembranes at higher capacities stay low but distinctly far from zero voltage and thus desirable to increase battery cell overall output voltage.

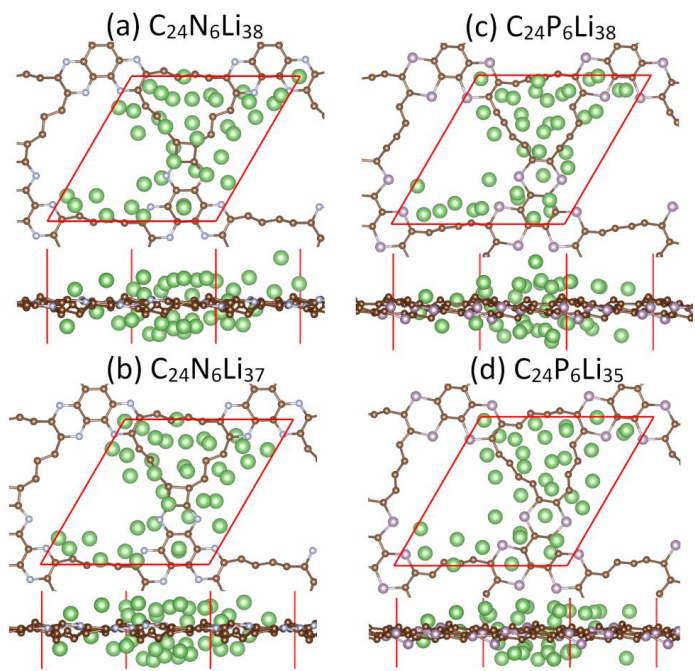

Fig. 9, Top and side views of energy minimized single-layer N-TpG and P-TpG with different content of Li adatoms.

In Fig. 9a and Fig. 9c, energy minimized N-TpG and P-TpG unit-cells with 38 adsorbed Li atoms are illustrated, respectively. According to these results, both nanosheets react flexibly upon the Li atoms adsorption, and no structural damage or bond breakage happen. Nevertheless, for the both monolayers some Li atoms were found to be pushed out of the first layer of covering Li atoms. Moreover, according to Bader charge analysis [80] results, some Li atoms could not transfer their electron charges to N-TpG or P-TpG monolayers. Our further simulations along with the subsequent Bader charge analysis confirmed that in the cases of energy minimized single-layer N-TpG and P-TpG unit-cells, with 37 and 35 Li atoms, respectively (as shown in Fig. 9), all adsorbed Li atoms were fully ionized (the overall efficiencies in the charge transfer to the substrates were found to be over 99%). According to these results, charge capacities of P-TpG and N-TpG for Li-ions storage are predicted to be as high as 1979 mAh/g and 2664 mAh/g, respectively. These charge capacities are outstandingly high, especially by taking into consideration that the charge capacities of commercial graphite and $TiO_2$ anode materials for Li-ions storage were reported to be 372 mAh/g [93] and 200 mAh/g [94], respectively. As pointed out earlier, our first-principles modelling results for B-graphdiyne [54], confirmed that B-graphdiyne can yield ultrahigh charge capacities for Na-, Ca- and Li-ion storage. This way, one can also expect that N-TpG and P-TpG nanosheets may also exhibit remarkably high storage capacities for Na-, K-, Mg- and Ca-ion storage, which can be an attractive topic for the future studies.

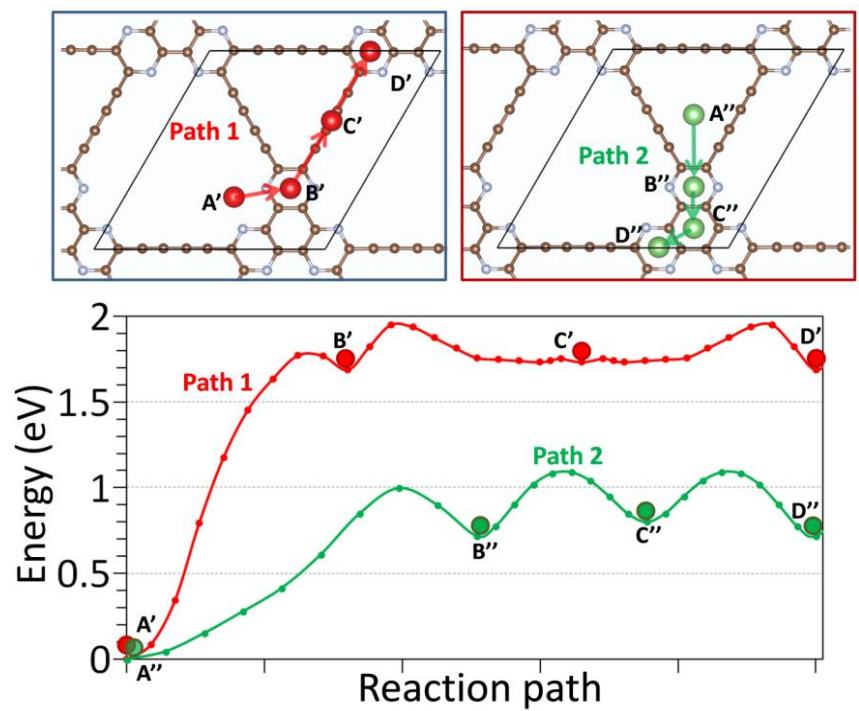

Fig. 10, Nudged-elastic band results for two different paths for a single Li atom diffusion over the single-layer N-TpG and the corresponding energy profiles.

Fast charging of Li-ion batteries, plays a critical role for many advanced technologies, like; electric vehicles and mobile communications. Since N-TpG nanosheet was predicted to yield the highest charge capacity, we next explore the ionic conductivity of this monolayer. To this aim, we employed the CiNEB method to simulate the diffusion of a single Li adatom and predict corresponding energy barriers. To this aim, as depicted in Fig. 10 two different paths for a single Li adatom diffusion were considered, representatives of Li-ions movements in this system. Path1 starts by hoping of the Li atom from the most stable adsorption site (see Fig, 8c) over the center of $C_4N_2$ hexagonal ring, and continues by diffusion over 6 membered carbon chains and ending on the facing $C_4N_2$ hexagonal ring hollow site. Path2 includes the hoping of the Li atom from the second most stable adsorption site (see Fig, 8d) over the center of $C_4N_2$ hexagonal ring, then jumping over the full carbon hexagonal ring and ending by hoping over equivalent $C_4N_2$ hexagonal ring hollow site. In Fig. 10 the corresponding relative energies for predicted images along the path1 and path2 are illustrated. As expected, for the hoping of the Li atom from the first and second most stable adsorption sites over the center of $C_4N_2$ hexagonal ring, high energy barriers of 1.8 eV and 1.0 eV has to be passed. Interestingly, for the diffusion of the Li atom from the center of $C_4N_2$ hexagonal ring, toward the hollow center of 6 membered carbon chain, only a considerably lower energy barrier of 0.37 eV has to be passed. On the other side, for the hoping of a single Li-ion over the hollow sites of hexagonal rings, the maximum energy barrier to overtake was measured to be 0.19 eV. These results confirm that the diffusion rate of Li-ions over the N-TpG is not a uniform process and depends strongly on the ions location and the direction of current (charging or discharge) as well. For the charging process in which the Li-ions diffuse over the carbon chains and hexagonal rings to fill the adsorption sites, the maximum energy barrier to be overtaken is 0.37 eV, which is very close to that over the graphene ~0.37 eV [95]. Moreover, porous structure of N-TpG provides excellent conditions for through-plane diffusion of Li-ions, whereas such a process for other densely packed 2D materials is considerably hard [96,97]. It can be therefore concluded that N-TpG anode material can be charged considerably faster, as compared with graphite anode material. Nevertheless, high energy barriers of 1.8 eV and 1.0 eV for the hopping of Li atoms form the in-plane most stable adsorption sites over the N-TpG monolayer surface suggest that at the last stages of the discharging process, the diffusion rates for Li-ions may slow-down and higher over-potentials might be required to drive the Li-ions.

Since path2 was found to dominate the energy barrier for the charging process, we also examined this path for the diffusion of a single Li atom over the TpG and P-TpG monolayers (results for the energy barriers are shown in Fig. S2 of SI document). It was found that for the both TpG and P-TpG monolayers, the maximum energy barriers for the hoping of the Li adatom over the hexagonal rings are ~0.42 eV, higher than that we found for the N-TpG (0.37 eV) nanosheet. These results reveal that in order to achieve fast charging rates, N-TpG is more promising in comparison with P-TpG and TpG nanosheets. For path2 nevertheless the energy barriers for the hoping of the Li atom from the most stable adsorption site over the neighbouring hexagonal ring, were estimated to be 0.9 eV and 0.75 eV for the TpG and P-TpG, respectively, which are lower than the energy barrier for the same path in the case of N-TpG (1.0 eV).

## 4. Concluding remarks

In a recent experimental advance, triphenylene-graphdiyne (TpG) nanomembranes were fabricated using a liquid/liquid interfacial method on the basis of hexaethynyltriphenylene chemical agent. Motivated by successful synthesis of TpG nanosheets, in this work we also predicted nitrogenated-, phosphorated- and arsenicated-TpG nanosheets, with similar atomic lattices as that of the pristine TpG. We then employed first-principles DFT calculations to obtain the energy minimized atomic lattices and elaborately explore mechanical properties, thermal stability, electronic and optical responses of free-standing and single-layer TpG, N-TpG, P-TpG and As-TpG. Mechanical/failure response of aforementioned monolayers were analyzed by performing uniaxial tensile simulations. Obtained results suggest convincingly isotropic elastic response of these monolayers. Mean elastic modulus of single-layer TpG, N-TpG, P-TpG and As-TpG were predicted to be 88 N/m, 93 N/m, 65 N/m and 56 N/m, respectively. It was found that the studied nanomembranes exhibit higher tensile strengths along the zigzag direction than the armchair direction. Tensile strengths of TpG, N-TpG, P-TpG and As-TpG along the zigzag direction were found to be 11.95 N/m, 10.37 N/m, 8.04 N/m and 6.34 N/m, respectively. Underlying mechanisms resulting in anisotropic tensile strengths and failure behaviour were also analyzed. AIMD simulations highlight outstanding thermal stability of studied monolayers, since their atomic lattices can stay intact at the high temperature of 2000 K. First-principles calculations confirm that single-layer TpG, As-TpG, P-TpG and N-TpG show direct band-gap semiconducting electronic characters, with band-gaps of 1.94 eV, 0.88 eV, 1.54 eV and 1.91 eV, respectively, according to the HSE06 method results. The first absorption peak for TpG and N-TpG nanosheet occur at visible range, 1.88 eV and 1.72 eV, respectively, while for As-TpG and P-TpG happen at 0.81 eV and 1.38 eV, respectively, which are in IR and NIR range of light, respectively. First absorption peaks reveal that these novel 2D nanostructures can absorb the visible, IR and NIR light, highlighting their promising prospect for applications in optoelectronics and nanoelectronics.

We particularly investigate the potential application of TpG, N-TpG, P-TpG and As-TpG nanomembranes as anode electrodes for Li-ions storage. According to DFT results, application prospects of TpG and As-TpG nanosheets as anode materials were found to be limited, because of very low voltage values and remarkable structural distortions, respectively. As a highly attractive finding, P-TpG and N-TpG nanomembranes were predicted to yield ultrahigh charge capacities of 1979 mAh/g and 2664 mAh/g, respectively, for Li-ions storage. Predicted open circuit voltage profiles reveal promising performance of P-TpG and N-TpG nanosheets. Climbing image nudged elastic band method results suggest that N-TpG nanomembranes can be a favourable anode material to conduct fast charging rates. The insights provided by the extensive first-principles results highlight the outstanding physics and chemistry of TpG based nanomembranes, and will hopefully motivate future experimental and theoretical studies. Likely to other members of graphdiyne family, TpG, N-TpG, P-TpG and As-TpG may also exhibit promising application prospects for hydrogen storage, catalysts and thermoelectricity.

**Acknowledgment**
B. M. and T. R. greatly acknowledge the financial support by European Research Council for COMBAT project (Grant number 615132).

**Data availability**
The energy minimized atomic lattices, HSE06 results and NEB predictions for the diffusion of a Li atom over the TpG and P-TpG are available to download.

# Supporting Information

## 1. Atomic structures of TpG, As-TpG, P-TpG and N-TpG unit-cells in VASP POSCAR

### 1.1 TpG ($C_{30}H_6$)

```
C30H6
   1.00000000000000
     13.7832044970202006    0.0000000000000000    0.0000000000000000
      6.8916025368744336   11.9366094101015108    0.0000000000000000
      0.0000000000000000    0.0000000000000000   20.0000000000000000
   C    H
    30    6
Direct
  0.1107941282860105  0.9977659620707078  0.5000000000000000
  0.2121823742690801  0.9990062476806315  0.5000000000000000
  0.3014873301163093  0.9988879331601408  0.5000000000000000
  0.3991159621055589  0.9987728686971017  0.5000000000000000
  0.4886310918323246  0.9982224456096817  0.5000000000000000
  0.5913252973583665  0.9969287256291931  0.5000000000000000
  0.7985808202942120  0.9963836624308872  0.5000000000000000
  0.6945398005436718  0.2062641625652585  0.5000000000000000
  0.9045971421994324  0.9965960238933097  0.5000000000000000
  0.9043836395002322  0.1004909328797717  0.5000000000000000
  0.0077089489060483  0.0977170595478501  0.5000000000000000
  0.5943466995487739  0.0969421432737066  0.5000000000000000
  0.6947455993093024  0.1000373918484389  0.5000000000000000
  0.5939771637204903  0.3096784669786505  0.5000000000000000
  0.7981740320662734  0.2065120877056981  0.5000000000000000
  0.5906339364763805  0.4128688980641231  0.5000000000000000
  0.4877927989591038  0.5141785396712674  0.5000000000000000
  0.3985102781971198  0.6033605243300736  0.5000000000000000
  0.3011177424497084  0.7011357408728364  0.5000000000000000
  0.2122864110872891  0.7907367903363612  0.5000000000000000
  0.1109699279395733  0.8934025377689839  0.5000000000000000
  0.7949241080460527  0.3101210309525098  0.5000000000000000
  0.6948360499071026  0.4130934439101941  0.5000000000000000
  0.6959938452036888  0.5145546890814697  0.5000000000000000
  0.6963131125884523  0.6036250987419947  0.5000000000000000
  0.6967449060829011  0.7009766793931897  0.5000000000000000
  0.6970114675360094  0.7900475943085542  0.5000000000000000
  0.6956788901766704  0.8927199355786666  0.5000000000000000
  0.7956685708978100  0.8958735856486334  0.5000000000000000
  0.0080620648544027  0.8962081824329421  0.5000000000000000
  0.0105969645846795  0.1751493551151413  0.5000000000000000
  0.8722585948586217  0.3131938693173137  0.5000000000000000
  0.5139660107496491  0.1742989576317830  0.5000000000000000
  0.5136967470347028  0.3124873702724059  0.5000000000000000
  0.8731081473495292  0.8155450585932166  0.5000000000000000
  0.0111260969644580  0.8158110360072569  0.5000000000000000
```

### 1.2 As-TpG ($C_{24}As_6$)

```
C24As6
1.00000000000000
    14.7667203200176793    0.0000000000000000    0.0000000000000000
     7.3833601600088370   12.7883549277480597    0.0000000000000000
     0.0000000000000000    0.0000000000000000   20.0000000000000000
   C    As
    24    6
Direct
  0.1364673385064663  0.9802776936070113  0.5000000000000000
  0.2307801365404998  0.9804346993628442  0.5000000000000000
  0.3137239140006717  0.9819669639932519  0.5000000000000000
```

```
       0.4043091100060995   0.9819669639932519   0.5000000000000000
       0.4887851670966341   0.9804346993628442   0.5000000000000000
       0.5832549328865188   0.9802776936070113   0.5000000000000000
       0.7987107911505823   0.0031783735855282   0.5000000000000000
       0.7032733549972022   0.1980179718199165   0.5000000000000000
       0.8981108832638892   0.0031783735855282   0.5000000000000000
       0.8980913174274027   0.0986327983675341   0.5000000000000000
       0.7032758572050459   0.0986327983675341   0.5000000000000000
       0.7987086901828891   0.1980179718199165   0.5000000000000000
       0.5833621399559539   0.4363897648614730   0.5000000000000000
       0.4888945436582172   0.5307441209321530   0.5000000000000000
       0.4043868376605069   0.6136681439382455   0.5000000000000000
       0.3138332570825698   0.7043247512290872   0.5000000000000000
       0.2309251152564452   0.7888609214781732   0.5000000000000000
       0.1365447464527579   0.8833291630617693   0.5000000000000000
       0.6802481431825588   0.4363897648614730   0.5000000000000000
       0.6803612944096115   0.5307441209321530   0.5000000000000000
       0.6819450064012709   0.6136681439382455   0.5000000000000000
       0.6818420396883356   0.7043247512290872   0.5000000000000000
       0.6802139512653692   0.7888609214781732   0.5000000000000000
       0.6801260194854803   0.8833291630617693   0.5000000000000000
       0.0197754985212765   0.1163240054924786   0.5000000000000000
       0.5639005239862358   0.1163240054924786   0.5000000000000000
       0.5639020424799973   0.3197980811002577   0.5000000000000000
       0.8162998944197494   0.3197980811002577   0.5000000000000000
       0.8162337278412397   0.8638892311702282   0.5000000000000000
       0.0198770469885568   0.8638892311702282   0.5000000000000000
```

### 1.3 P-TpG (C$_{24}$P$_6$)

```
C24P6
1.00000000000000
    14.4146613465835092    0.0000000000000000    0.0000000000000000
     7.2073306732917573   12.4834629131230308    0.0000000000000000
     0.0000000000000000    0.0000000000000000   20.0000000000000000
   C    P
   24    6
Direct
  0.1276412402100209   0.9860583397585486   0.5000000000000000
  0.2252446452506946   0.9853399887104484   0.5000000000000000
  0.3103690668486241   0.9863251069914990   0.5000000000000000
  0.4033058141598788   0.9863251069914990   0.5000000000000000
  0.4894153690388636   0.9853399887104484   0.5000000000000000
  0.5863003850314197   0.9860583397585486   0.5000000000000000
  0.7985925367430041   0.0002411569888881   0.5000000000000000
  0.7002418041256420   0.2011646326467391   0.5000000000000000
  0.9011663542681215   0.0002411569888881   0.5000000000000000
  0.9011732364665735   0.0985903082154993   0.5000000000000000
  0.7002364283179241   0.0985903082154993   0.5000000000000000
  0.7985935802276267   0.2011646326467391   0.5000000000000000
  0.5862983461835949   0.4276317552895316   0.5000000000000000
  0.4894207769739280   0.5252403558123943   0.5000000000000000
  0.4032872755202508   0.6103281589158704   0.5000000000000000
  0.3103497873566116   0.7032647171847070   0.5000000000000000
  0.2252531380569286   0.7893976458072057   0.5000000000000000
  0.1276510016192805   0.8862744130004115   0.5000000000000000
  0.6860699465268592   0.4276317552895316   0.5000000000000000
  0.6853388262136882   0.5252403558123943   0.5000000000000000
  0.6863845535639016   0.6103281589158704   0.5000000000000000
  0.6863855434586528   0.7032647171847070   0.5000000000000000
  0.6853492041358461   0.7893976458072057   0.5000000000000000
  0.6860745143803298   0.8862744130004115   0.5000000000000000
  0.0161040253691326   0.1147625526571291   0.5000000000000000
  0.5691334499737223   0.1147625526571291   0.5000000000000000
  0.5691304603486221   0.3160941818557278   0.5000000000000000
  0.8147753757956403   0.3160941818557278   0.5000000000000000
```

```
  0.8147763164037912   0.8691233701654156   0.5000000000000000
  0.0161003194307751   0.8691233701654156   0.5000000000000000
```

**1.4 N-TpG (C$_{24}$N$_6$)**

```
C24N6
1.00000000000000
    13.4959505871951997    0.0000000000000000    0.0000000000000000
     6.7479752935976069   11.6878360567406396    0.0000000000000000
     0.0000000000000000    0.0000000000000000   20.0000000000000000
   C    N
   24    6
Direct
  0.1039224965222800   0.0017216691174582   0.5000000000000000
  0.2073247183131031   0.0036972738172949   0.5000000000000000
  0.2980264924387939   0.0044201510691551   0.5000000000000000
  0.3975533444920529   0.0044201510691551   0.5000000000000000
  0.4889780108696087   0.0036972738172949   0.5000000000000000
  0.5943557993602511   0.0017216691174582   0.5000000000000000
  0.7990985143578229   0.9934161602246405   0.5000000000000000
  0.6934271267001080   0.2074673416743877   0.5000000000000000
  0.9074853734175505   0.9934161602246405   0.5000000000000000
  0.9074903593424608   0.0990942370168924   0.5000000000000000
  0.6934153766406439   0.0990942370168924   0.5000000000000000
  0.7991055486255050   0.2074673416743877   0.5000000000000000
  0.5943609579273925   0.4038959587126243   0.5000000000000000
  0.4889875169798827   0.5073074036531509   0.5000000000000000
  0.3975670843976928   0.5980125945865851   0.5000000000000000
  0.2980279926793159   0.6975354961876121   0.5000000000000000
  0.2073336007753923   0.7889658702839455   0.5000000000000000
  0.1039214205480192   0.8943379639851443   0.5000000000000000
  0.7017431313599829   0.4038959587126243   0.5000000000000000
  0.7037050383669552   0.5073074036531509   0.5000000000000000
  0.7044203090157239   0.5980125945865851   0.5000000000000000
  0.7044365591330717   0.6975354961876121   0.5000000000000000
  0.7037005169406498   0.7889658702839455   0.5000000000000000
  0.7017405444668370   0.8943379639851443   0.5000000000000000
  0.0055238428208995   0.1017694895278183   0.5000000000000000
  0.5927066956512661   0.1017694895278183   0.5000000000000000
  0.5927122884907448   0.3054986799641668   0.5000000000000000
  0.8017890495450928   0.3054986799641668   0.5000000000000000
  0.8017860136199673   0.8926963941791541   0.5000000000000000
  0.0055175982008890   0.8926963941791541   0.5000000000000000
```

**2. HSE06 results for the EDOS of TpG, As-TpG, P-TpG and N-TpG monolayers.**

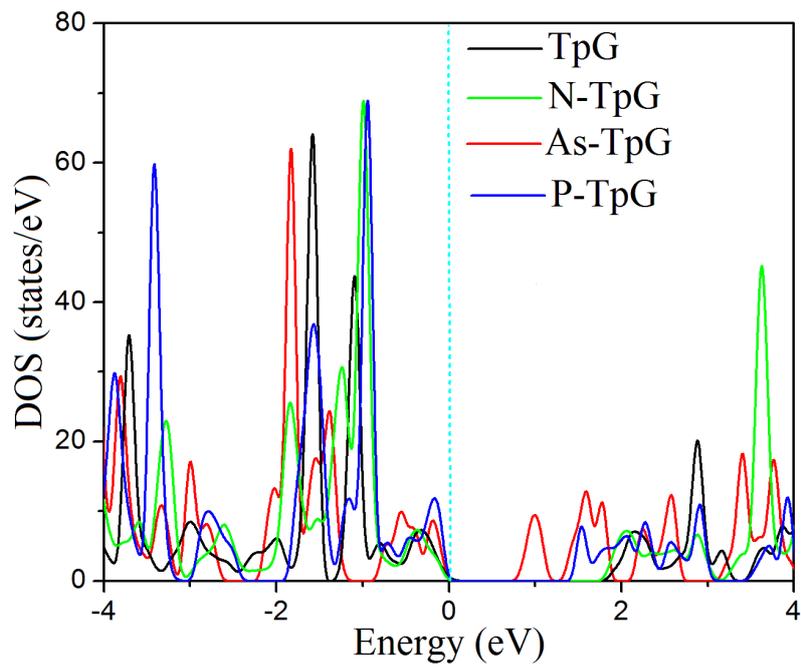

Fig. S1, Total EDOS of single-layer TpG, As-TpG, N-TpG and P-TpG predicted by the HSE06 approach. The Fermi energy is aligned to zero.

**3. Nudged-elastic band results for the diffusion of a single Li atom over the N-TpG, P-TpG and TpG along the path2.**

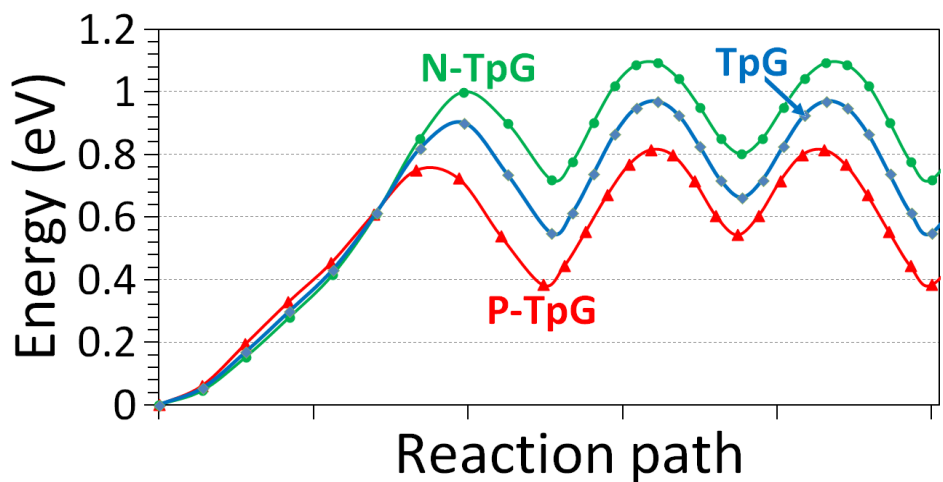

Fig. S2, Nudged-elastic band results for a single Li atom diffusion along the path2 (as depicted in Fig. 10) over the single-layer N-TpG, P-TpG and TpG.